\documentclass[preprint,aps,prd,groupedaddress,showpacs,floatfix,nofootinbib]
              {revtex4}%
\usepackage{graphicx}
\usepackage{amsmath}
\usepackage{bm}

\begin{document}

\preprint{ANL-HEP-PR-06-41}
\title{
Potential-model calculation of an order-$\bm{v}^2$ \\
nonrelativistic QCD matrix element 
}

\author{Geoffrey T.~Bodwin}
\affiliation{
High Energy Physics Division, 
Argonne National Laboratory, 
9700 South Cass Avenue, Argonne, Illinois 60439}

\author{Daekyoung Kang}
\affiliation{
Department of Physics, Korea University, Seoul 136-701, Korea}
\author{Jungil Lee}
\altaffiliation{Visiting faculty, Physics Department, Ohio State University,
Columbus, Ohio 43210, USA.}
\affiliation{
Department of Physics, Korea University, Seoul 136-701, Korea}


\date{\today}
\begin{abstract}
We present two methods for computing dimensionally-regulated NRQCD
heavy-quarkonium matrix elements that are related to the second
derivative of the heavy-quarkonium wave function at the origin. The
first method makes use of a hard-cutoff regulator as an intermediate
step and requires knowledge only of the heavy-quarkonium wave function.
It involves a significant cancellation that is an obstacle to
achieving high numerical accuracy. The second method is more direct
and yields a result that is identical to the Gremm-Kapustin relation,
but it is limited to use in potential models. It can be generalized to 
the computation of matrix elements of higher order in the heavy-quark 
velocity and can be used to resum the contributions to decay and 
production rates that are associated with those matrix elements.
We apply these methods to the Cornell potential model and compute a
matrix element for the $J/\psi$ state that appears in the leading
relativistic correction to the production and decay of that state
through the color-singlet quark-antiquark channel.
\end{abstract}

\pacs{12.38.-t, 12.39.Pn, 12.38.Bx, 13.20.Gd, 14.40.Gx}

\maketitle

\section{Introduction\label{intro}}

In the effective field theory nonrelativistic quantum chromodynamics
(NRQCD), the leading relativistic corrections to $S$-wave
heavy-quarkonium decay and production processes in the color-singlet
quark-antiquark channel are proportional to matrix elements that
are related to the second derivative of the quarkonium wave function at
the origin. These matrix elements are inherently nonperturbative in
nature. Their importance in phenomenological calculations has led to a
number of attempts to determine their values.

Even before the introduction of the NRQCD approach for quarkonium decay
and production \cite{Bodwin:1994jh}, these matrix elements appeared in
phenomenological studies of charmonium decays \cite{Kwong:1987ak}. Owing
to uncertainties that arise from the uncertainty in the charm-quark
mass $m_c$ and from uncalculated terms of higher order in the quantum
chromodynamic (QCD) strong coupling $\alpha_s$, such phenomenological
determinations have not led to accurate values for the matrix elements.
There also have been attempts to determine the matrix elements in
lattice calculations \cite{bks}. In this case, large uncertainties arise
because there is a substantial cancellation that occurs when one
converts from lattice to continuum dimensionally-regulated matrix
elements. In principle, one can determine these matrix elements
approximately by making use of the Gremm-Kapustin relation
\cite{Gremm:1997dq}, which expresses the matrix elements in terms of the
quarkonium and heavy-quark masses. (See Ref.~\cite{Braaten:2002fi} for
an example of this approach.) Unfortunately, this method is plagued by
large uncertainties in $m_c$. In both the lattice and
Gremm-Kapustin approaches, the uncertainties are so large that even the
signs of the matrix elements are in some doubt.

A further difficulty that complicates the calculation of the matrix
elements that are related to the second derivative of the wave function
at the origin is that they contain a linear ultraviolet (UV) divergence,
and, hence, must be regulated. Dimensional regularization of these
matrix elements is particularly useful because it is consistent with
existing calculations of quarkonium decay and production rates at
relative order $\alpha_s$ and $\alpha_s^2$.

In this paper, we present two methods for calculating the second
derivative of the wave function at the origin. In the first method, we
initially use a hard-cutoff regulator, with cutoff $\Lambda$, to define
the relevant matrix element. Then we compute the difference between the
hard-cutoff regularization and dimensional regularization in
perturbation theory. We subtract this difference from the hard-cutoff
matrix element. There remains a dependence of the matrix element on
$\Lambda$ that falls as $1/\Lambda$ in the limit $\Lambda\to\infty$.
That dependence can be removed by calculating at a number of values of
$\Lambda$ and extrapolating to $\Lambda=\infty$. However, in this
method, it is difficult to achieve high numerical accuracy because
large, cancelling $\Lambda$-dependent contributions appear. The second
method that we present bypasses the hard-cutoff step, but it is
applicable only to potential models for the wave function. It yields a
result that is identical to the Gremm-Kapustin relation
\cite{Gremm:1997dq} for the potential model. That result can be used to
resum certain contributions of higher order in $v$ to amplitudes that
are computed in NRQCD.

Having established a formal procedure for computing the relevant matrix
elements, we carry out an explicit computation for the $J/\psi$ (or
$\eta_c$) states in the Cornell potential model. We do not distinguish
between the $J/\psi$ and $\eta_c$ wave functions, which differ only in
corrections of relative order $v^2$, where $v$ is the velocity of the
heavy quark or antiquark in the quarkonium. ($v^2\approx 0.3$ in
charmonium and $v^2\approx 0.1$ in bottomonium.) In principle, if we
know the static heavy-quark-antiquark ($Q\bar Q$) potential exactly,
then we can calculate the quarkonium wave function of the leading $Q\bar
Q$ quarkonium Fock state up to corrections of relative order $v^2$.
Existing lattice data for the static $Q\bar Q$ potential yield values
for the string tension. We examine values for the parameters in the
Cornell potential that bracket the lattice values for the string
tension.

In our numerical calculations, the results from our two approaches agree
well and give a value for the second derivative of the wave function at
the origin that is in agreement with expectations from the $v$-scaling
rules of NRQCD \cite{Bodwin:1994jh}. The largest uncertainties in our
calculation are of relative order $v^2$. Therefore, our
determination of the second derivative of the wave function at the
origin is the most accurate to date and should be useful for
phenomenological studies of quarkonium production and decay.

The remainder of this paper is organized as follows: In
Sec.~\ref{model}, we give a brief description of the Cornell potential
model. In Sec.~\ref{matrix-els}, we discuss the NRQCD matrix elements
that are relevant to this work. Section~\ref{cutoff} contains a description
of the hard-cutoff regulator. In Sec.~\ref{difference}, we explain how
we compute the difference between hard-cutoff regularization and
dimensional regularization. The direct method of calculation, which
bypasses the hard-cutoff step, is discussed in Sec.~\ref{direct}, along
with its application to resummation of contributions of higher order in
$v$. In Sec.~\ref{decomposition}, we decompose the difference 
between a hard-cutoff matrix element  and a dimensionally-regulated 
matrix element into sums of short-distance coefficients times 
dimensionally-regulated matrix elements. Such a decomposition is used 
in existing calculations of the difference between a lattice matrix element 
and a dimensionally-regulated matrix element.
Sec.~\ref{numerical} contains our numerical
results and a discussion of them. We summarize our results in
Sec.~\ref{conclusions}. In Appendix~\ref{coulomb}, we illustrate our
methods for the case of a pure Coulomb potential, and, in
Appendix~\ref{integrals}, we list some integrals that are useful in our
analyses.

\section{Potential model\label{model}}
We compute matrix elements for the $J/\psi$ or $\eta_c$ states using a
potential model. In this model, we neglect the effects of the heavy-quark
spin, which are suppressed as $v^2$. Therefore, we do not distinguish
between the $J/\psi$ and $\eta_c$ wave functions or matrix elements. We
note that, if we knew the heavy-quark potential exactly, then we could
calculate the heavy-quarkonium wave function in a potential model up to
corrections of relative order $v^2$ (Ref.~\cite{Brambilla:1999xf}). We
make use of the Cornell potential model of Ref.~\cite{Eichten:1978tg}.
For appropriate choices of parameters, the Cornell potential
provides a reasonably good fit to heavy-quark potentials that are
measured in lattice calculations.\footnote{For a recent review that
discusses heavy-quark potentials from lattice measurements, see
Ref.~\cite{Bali:2000gf}.}

Now we summarize the methods that we use to constrain the parameters of
the Cornell potential and to solve the Schr\"odinger equation. We refer
the reader to Ref.~\cite{bkl} for further details.

The Cornell potential \cite{Eichten:1978tg} is given by
\begin{equation}
V(r)=-\frac{\kappa}{r}+\frac{r}{a^2},
\label{model-V}
\end{equation}
where the parameters $\kappa$ and $a$ determine the strength of 
Coulomb and linear potentials, respectively.
For a color-singlet $Q\bar{Q}$ pair, the Coulomb-strength
parameter $\kappa$ can be expressed in terms of an effective
strong coupling $\alpha_s$ as
\begin{equation}
\kappa=\alpha_s C_F,
\label{kappa}
\end{equation}
where $C_F=4/3$.
The parameter $a$ is related to the string tension $\sigma$ as
\begin{equation}
\sigma=\frac{1}{a^2}.
\label{sigma-par}
\end{equation}

Following Ref.~\cite{Eichten:1978tg}, we replace $\kappa$ and the
binding energy $\epsilon_{\textrm{B}}$ by dimensionless parameters
$\lambda$ and $\zeta$:
\begin{subequations}
\label{lambda-zeta}
\begin{eqnarray}
\kappa&=&(ma)^{-\frac{2}{3}}\,\lambda,\label{lambda}\\
\epsilon_{\textrm{B}}&=& m (ma)^{-\frac{4}{3}}\zeta.\label{zeta}
\end{eqnarray}
\end{subequations}
For a given value of $\lambda$, we fix the heavy-quark mass $m$ and the
parameter $a$ in the Cornell potential as follows. First, we require
that the energies that result from the solutions to the Schr\"odinger
equation match the measured values of the difference of the $J/\psi$ and
$\psi(2S)$ masses. We use $M_{J/\psi}=3.096916$~GeV and
$M_{\psi(2S)}=3.686093$~GeV. Second, we require that the wave function
at the origin matches a value that is derived from the measured value of
the leptonic width of the $J/\psi$ and the perturbative formula
\begin{equation}
\Gamma[J/\psi\to \ell^+\ell^-]
=\frac{4\pi e_c^2 \alpha^2}{m_c^2}|\psi(0)|^2
\left(1-\frac{8}{3}\frac{\alpha_s}{\pi}\right)^2.
\label{lept-width}
\end{equation}
Here, $\psi(0)$ is the wave function at the origin and $e_c=2/3$ is the
fractional electric charge of the charm quark. 
In Ref.~\cite{Braaten:2002fi}, $\psi(0)$ is estimated by using the 
the formula (\ref{lept-width}) at both leading order in $\alpha_s$ (LO) 
and next-to-leading order in $\alpha_s$ (NLO). The results are
\begin{equation}                                                         
\psi(0)=                 
\left\{                                                                 
\begin{array}{lcl}
0.18619~\textrm{GeV}^{3/2}&& \textrm{(LO)}
\\
0.23629~\textrm{GeV}^{3/2}&& \textrm{(NLO)},
\end{array}
\right.
\label{psi0-value}
\end{equation}
where, for convenience, we have taken $\psi(0)$ to be positive and real.
In order to estimate the effects of the uncertainty in $\psi(0)$, we
carry out our calculations for both the LO and NLO values of $\psi(0)$.
For a more detailed discussion of the determination of $m$ and $a$, see 
Ref.~\cite{bkl}.

Values of $m$, $a$, and the scaled energies $\zeta$ of the $1S$ and $2S$
states for various values of $\lambda$ are shown in
Tables~\ref{table:ma-lo} and \ref{table:ma-nlo}, along with values for
$\alpha_s$ from Eq.~(\ref{kappa}), $\sigma$ from Eq.~(\ref{sigma-par}),
and $\gamma_{\textrm{C}}$ from Eq.~(\ref{gamma-C}). 
Table~\ref{table:ma-lo} contains values of the potential-model
parameters that correspond to the LO value of $\psi(0)$, while
Table~\ref{table:ma-nlo} contains those that correspond to the NLO value
of $\psi(0)$.
 \begin{table}
 \caption{\label{table:ma-lo}
Potential-model parameters and derived quantities as a function of the
strength $\lambda$ of the Coulomb potential. The definitions of the
parameters and derived quantities are given in the text. The parameters
are computed using the inputs $M_{J/\psi}=3.096916$~GeV, 
$M_{\psi(2S)}=3.686093$~GeV, and the LO value 
$\psi(0)=0.18619$~GeV$^{3/2}$, as is described in the text.
}
 \begin{ruledtabular}
 \begin{tabular}{c|lllllllll}
$\lambda$& 
0& 0.2     & 0.4     & 0.6     & 0.7     & 0.8    & 1.0   & 1.2     & 1.4\\
\hline
$\zeta_{10}$&
2.33811& 2.16732& 1.98850& 1.80107& 1.70394& 1.60441& 1.39788& 1.18084& 0.95264
\\
$\zeta_{20}$&
4.08790& 3.97017& 3.85003& 3.72747& 3.66528& 3.60249& 3.47510& 3.34529& 3.21307
\\
$m$~(GeV)&
1.70670& 1.51548& 1.35120& 1.21003& 1.14710& 1.08877& 0.98458& 0.89501& 0.81796
\\
$a$~(GeV$^{-1}$)&
1.97932& 2.08520& 2.19805& 2.31833& 2.38139& 2.44648& 2.58295& 2.72816& 2.88253
\\
$\sqrt{\sigma}$~(GeV)&
0.50522& 0.47957& 0.45495& 0.43134& 0.41992& 0.40875& 0.38715& 0.36655& 0.34692
\\
$\sigma$~(GeV$^2$)&
0.25525& 0.22999& 0.20698& 0.18606& 0.17633& 0.16708& 0.14989& 0.13436& 0.12035
\\
$\alpha_s$&
0.     & 0.06966& 0.14519& 0.22624& 0.26866& 0.31225& 0.40255& 0.49634& 0.59272
\\
$\gamma_{\textrm{C}}$~(GeV)&
0.     & 0.07037& 0.13079& 0.18250& 0.20545& 0.22664& 0.26423& 0.29615& 0.32322
 \end{tabular}
 \end{ruledtabular}
 \end{table}

 \begin{table}
 \caption{\label{table:ma-nlo} As in Table~\ref{table:ma-lo}, except
that the potential-model parameters are computed using the NLO value
$\psi(0)=0.23629$~GeV$^{3/2}$.
}
 \begin{ruledtabular}
 \begin{tabular}{c| lll lll lll}
$\lambda$& 
0.2    & 0.4    & 0.6    & 0.8    & 0.9    & 1.0    & 1.1    & 1.2    &  1.4    \\
\hline
$\zeta_{10}$&
2.16732& 1.98850& 1.80107& 1.60441& 1.50242& 1.39788& 1.29071& 1.18084&  0.95264
\\
$\zeta_{20}$&
3.97017& 3.85003& 3.72747& 3.60249& 3.53910& 3.47510& 3.41050& 3.34529&  3.21307
\\
$m$~(GeV)&
2.08228& 1.85655& 1.66259& 1.49597& 1.42168& 1.35282& 1.28896& 1.22975&  1.12388
\\
$a$~(GeV$^{-1}$)&
1.92598& 2.03021& 2.14130& 2.25967& 2.32171& 2.38571& 2.45174& 2.51984&  2.66243
\\
$\sqrt{\sigma}$~(GeV)&
0.51922& 0.49256& 0.46700& 0.44254& 0.43072& 0.41916& 0.40787& 0.39685&  0.37560
\\
$\sigma$~(GeV$^2$)&
0.26959& 0.24262& 0.21809& 0.19584& 0.18552& 0.17570& 0.16636& 0.15749&  0.14107
\\
$\alpha_s$&
0.05942& 0.12387& 0.19301& 0.26638& 0.30448& 0.34342& 0.38310& 0.42343&  0.50566
\\
$\gamma_{\textrm{C}}$~(GeV)&
0.08249& 0.15331& 0.21393& 0.26567& 0.28859& 0.30972& 0.32920& 0.34714&  0.37887
 \end{tabular}
 \end{ruledtabular}
 \end{table}


Lattice measurements of the heavy-quark potential yield values for
effective coupling $\alpha_s$ of 0.22 in the quenched case and
approximately 0.26 in the unquenched case \cite{Bali:2000gf}. A lattice
measurement of the string tension $K=\sigma$ (Ref.~\cite{Booth:1992bm})
gives $Ka_{\rm L}^2=0.0114(2)$ at a lattice coupling $\beta=6.5$, where 
$a_{\rm L}$ is
the lattice spacing. Lattice calculations of the hadron spectrum at
$\beta=6.5$ yield values for $1/a_{\rm L}$ of $3.962(127)$~GeV
(Refs.~\cite{Gupta:1996sa,Kim:1993gc}) and $3.811(59)$~GeV
(Refs.~\cite{Gupta:1996sa,Kim:1996cz}). These yield values of the string
tension of $K=0.1790\pm 0.0119$ and $K=0.1656\pm 0.0059$, respectively.

Comparing the results of these lattice measurements with the LO
parameters in Table~\ref{table:ma-lo}, we see that the values of the
string tension at $\lambda=0.7$ and $0.8$ span the range of lattice
results for the string tension, while the values of $\alpha_s$ at
$\lambda=0.6$ and $0.7$ span the range of lattice results for $\alpha_s$.
Comparing the results of the lattice measurements with the NLO
parameters in Table~\ref{table:ma-nlo}, we see that the values of
the string tension at $\lambda=1.0$ and $1.1$ span the range of lattice
results for the string tension. However, the values of $\alpha_s$ at
$\lambda=0.9$, $1.0$, and $1.1$ are all larger than the lattice values.
It is not clear whether this discrepancy between the lattice and NLO
potential-model values for $\alpha_s$ arises from the use of an
inaccurate value for $\psi(0)$, from effects due to the running of
$\alpha_s$, which are not taken into account in the fits to the lattice
data, or from the absence of corrections of relative order $v^2$ in
the Cornell potential model. However, we note that the NLO values for
$\alpha_s$ at $\lambda=0.9$, $1.0$, and $1.1$ do not differ greatly from
the value of the running $\alpha_s$ at the scale of the heavy-quark
momentum $m_cv$.

Finally, we mention that we obtain the $J/\psi$ wave function by expressing
the radial part of the Schr\"odinger equation as a difference equation,
which we integrate numerically. See Ref.~\cite{bkl} for details.

\section{ NRQCD matrix elements \label{matrix-els}}
In this section we describe the NRQCD matrix elements that are relevant 
to our calculation.

In the rest frame of an $S-$wave heavy quarkonium $H$ in a spin-singlet
(${}^1S_0$) or spin-triplet (${}^3S_1$) state, one can express the wave
function at the origin of the leading $Q\bar Q$ Fock state in terms of
the following color-singlet NRQCD matrix elements
\cite{Bodwin:1994jh}:
\begin{subequations}
\label{psi0}
\begin{eqnarray}
\psi(0)&=&
\int\frac{d^3p}{(2\pi)^3}
\widetilde{\psi}(\bm{p})
=\frac{1}{\sqrt{2N_c}}
\langle 0|\chi^\dagger\psi|H({}^1S_0)\rangle,
\\
\bm{\epsilon}\psi(0)&=&\bm{\epsilon}\int\frac{d^3p}{(2\pi)^3}
\widetilde{\psi}(\bm{p})
=\frac{1}{\sqrt{2N_c}}
\langle 0|\chi^\dagger\bm{\sigma}\psi|H({}^3S_1)\rangle.
\end{eqnarray}
\end{subequations}
Here $\psi$ and $\chi^\dagger$ are Pauli spinor fields
that annihilate a quark and an antiquark, respectively.
The bilinear operators involving $\psi^\dagger$ and $\chi$ are evaluated
at zero space-time position. $\bm{\sigma}$ is a Pauli matrix, and
$\bm{\epsilon}$ is the quarkonium polarization vector.
$\widetilde{\psi}(\bm{p})$ is the momentum-space wave function for the
leading $Q(\bm{p})\bar{Q}(-\bm{p})$ Fock state in the quarkonium. 
The wave function is, of course, gauge dependent. 
Throughout this paper, we work in the Coulomb gauge.
The normalization factor $1/\sqrt{2N_c}$ accounts for the traces in the
SU(2)-spin and SU(3)-color spaces. In equating the wave functions for the
spin-singlet and spin-triplet cases, we are ignoring effects of 
relative order $v^2$.

Relativistic corrections to the production and decay rates for
a heavy quarkonium involve matrix elements that are related to the 
second derivative of the wave function at the origin:
\begin{subequations}
\label{psi20}
\begin{eqnarray}
\psi^{(2)}(0)&\equiv&
\int\frac{d^3p}{(2\pi)^3}
\bm{p}^2 
\widetilde{\psi}(\bm{p})
=\frac{1}{\sqrt{2N_c}}
\langle 0|\chi^\dagger\left(-\bm{\nabla}^2\right)\psi|H({}^1S_0)\rangle,
\\
\bm{\epsilon}\psi^{(2)}(0)&\equiv&\bm{\epsilon}\int\frac{d^3p}{(2\pi)^3}
\bm{p}^2
\widetilde{\psi}(\bm{p})
=\frac{1}{\sqrt{2N_c}}
\langle 0|\chi^\dagger\bm{\sigma}\left(-\bm{\nabla}^2\right)
\psi|H({}^3S_1)\rangle.
\end{eqnarray}
\end{subequations}
Usually, these operator matrix elements are written in terms of the
covariant derivative $\bm{D}$ (Ref.~\cite{Bodwin:1994jh}), rather than
$\bm{\nabla}$. However, in the Coulomb gauge, the difference between the
$\bm{D}$ and $\bm{\nabla}$ is suppressed as $v$
(Ref.~\cite{Bodwin:1994jh}). 

The quantity $\psi^{(2)}(0)$ is the focus of this paper. It is 
common in phenomenology to make use of a parameter 
\begin{equation}
\langle v^2\rangle=\frac{\psi^{(2)}(0)}{m_c^2\psi(0)},
\label{v-sq}
\end{equation}
where $m_c$ is the charm-quark pole mass, which we distinguish from the 
parameter $m$ that appears in our potential model.
Note that $\psi^{(2)}(0)$ is different from the expectation value of 
$\bm{p}^{2}$:
\begin{equation}
\psi^{(2)}(0)\neq
\int\frac{d^3p}{(2\pi)^3}
\,\bm{p}^{2}
\widetilde{\psi}^*(\bm{p})
\widetilde{\psi}(\bm{p}).
\end{equation}

Let us investigate the ultraviolet behavior of the matrix elements in
Eq.~(\ref{psi20}). At large momentum $|\bm{p}|$, the interaction of the
$Q\bar{Q}$ pair in QCD is dominated by the Coulomb potential. In this
limit, the bound-state wave function approaches the pure Coulomb wave
function:
\begin{equation}
\widetilde{\psi}(\bm{p})
\sim \frac{1}{(\bm{p}^2+\gamma^2_{\textrm{C}})^2}.
\label{coulomb-wf}
\end{equation}
Here $\gamma_{\textrm{C}}$ is a parameter that is related to the binding
energy of Coulomb interaction:
\begin{equation}
\gamma_{\textrm{C}}=\frac{1}{2}\,\alpha_s C_F m,
\label{gamma-C}
\end{equation}
where $\alpha_s$ is the effective strong coupling, $C_F=4/3$, and 
$m$ is the quark mass. 
Substituting Eq.~(\ref{coulomb-wf}) into Eq.~(\ref{psi20}), 
we see that the matrix elements in Eq.~(\ref{psi20}) have a linear ultraviolet 
divergence. Hence, in order to define them, we must impose a regulator.

\section{Hard-cutoff regulator\label{cutoff}}

Our ultimate goal is to regulate the matrix elements in 
Eq.~(\ref{psi20}) using dimensional regularization. However, as an 
intermediate step, we impose a hard-cutoff regulator. In principle, the 
methods that we use could be employed in lattice calculations of the 
matrix elements, in which the lattice supplies the hard-cutoff 
regulator. However, for the purposes of the present work, we make use of 
a simple, analytic UV regulator in momentum space:
\begin{equation}
\label{psi-lam}
\psi^{(2)}_\Lambda(0)=
\int\frac{d^3p}{(2\pi)^3}\frac{\Lambda^2}{\bm{p}^2+\Lambda^2}\, \bm{p}^2
\widetilde{\psi}(\bm{p}),
\end{equation}
where $\Lambda$ is the cutoff. 

We introduce the Fourier transform of the wave function to coordinate 
space:
\begin{equation}
\psi(\bm{x})=\int\frac{d^3{p}}{(2\pi)^3}e^{i\bm{p}\cdot\bm{x}} 
\widetilde{\psi}(\bm{p}).
\label{fourier-tx-wavefn}
\end{equation}
For $S$-wave states, we can write $\psi(\bm{x})=R(r)/\sqrt{4\pi}$, where
$r=|\bm{x}|$ and $R(r)$ is the radial wave function. Substituting
Eq.~(\ref{fourier-tx-wavefn}) into  Eq.~(\ref{psi-lam}), we can carry out
the angular integration over $\bm{p}$ to obtain an expression in
coordinate space:
\begin{equation}
\label{psi-lamr}
\psi^{(2)}_\Lambda(0)=
\frac{\Lambda^2}{\sqrt{4\pi}}
\left[R(0)-\Lambda^2\int_0^\infty rR(r)e^{-\Lambda r}dr
\right].
\end{equation}

\section{Difference between hard-cutoff and dimensional 
regularization\label{difference}}
Now we work out the difference between the hard-cutoff matrix 
element and the dimensionally-regulated matrix element
\begin{equation}
\label{dpsi-lam}
\Delta\psi^{(2)}(0)\equiv
\psi_{\Lambda}^{(2)}(0) - \psi_{\textrm{DR,}\Lambda}^{(2)}(0).
\end{equation}
Because $\Delta\psi^{(2)}(0)$ involves a change in the ultraviolet
cutoff, it is sensitive only to the high-momentum part of the $\bm{p}$
integration in the momentum-space definition of $\psi^{(2)}(0)$ in
Eq.~(\ref{psi20}). Therefore, we can compute $\Delta\psi^{(2)}(0)$ in
perturbation theory. Here, we carry out the computation in lowest
(one-loop) order.\footnote{One could, in principle, compute
corrections of higher order in $\alpha_s$. Ultimately, the series of
these corrections diverges, owing to the renormalon ambiguity that
appears in dimensionally-regulated matrix elements. That ambiguity is
canceled by a corresponding ambiguity in the NRQCD short-distance
coefficients, provided that one computes to the same loop order in both
the matrix elements and the short-distance coefficients. See
Ref.~\cite{Bodwin:1998mn} for a discussion of this point.} We note, 
however, that the perturbative calculation of $\Delta\psi^{(2)}(0)$ does 
not contain all of the corrections that behave as powers of $1/\Lambda$
to the difference between $\psi_{\Lambda}^{(2)}(0)$ and 
$\psi_{\textrm{DR,}\Lambda}^{(2)}(0)$. Therefore, it is necessary to 
take the limit $\Lambda\to\infty$ in order to remove the effects of such 
power corrections.

\begin{figure}
\includegraphics[width=7cm]{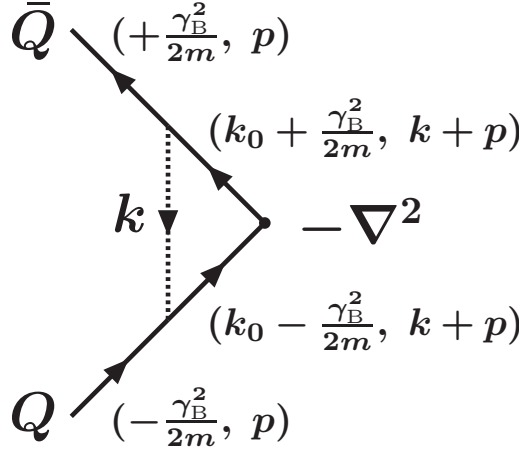}
\caption{Feynman diagram corresponding to $\tilde{I}^{(2)}(\bm{p})$, 
which is the one-loop renormalization of the operators in 
Eq.~(\ref{psi20}). The solid lines represent the heavy quark and 
antiquark, and the dotted line represents the potential between them.}
\label{fig:Feynman}
\end{figure}

We begin by using the Bethe-Salpeter equation (or, equivalently, the
Schr\"odinger equation) to expose one explicit loop in the wave
function. Then, we have
\begin{equation}
\label{1-loop-renorm}
\psi^{(2)}(0)=
\int\frac{d^3p}{(2\pi)^3}\, 
\tilde{I}^{(2)}(\bm{p}) 
\,\widetilde{\psi}(\bm{p}),
\end{equation}
where $\tilde{I}^{(2)}(\bm{p})$ is the quantity that is 
represented by the Feynman diagram in Fig.~\ref{fig:Feynman}.
In Fig.~\ref{fig:Feynman}, the dotted line represents the potential.
Since the loop integral is dominated by large momenta, we can
approximate the potential by the Coulomb part. 

We emphasize that we can always make this approximation in QCD,
irrespective of the use of a potential model. Asymptotic freedom allows
one to evaluate the high-momentum loop in Fig.~\ref{fig:Feynman} in
perturbation theory. Then, in the Coulomb gauge, the Coulomb-gluon
interactions with the heavy quark and antiquark give the result of
leading order in $v$, while the transverse-gluon interactions are
suppressed as $v^2$ (Ref.~\cite{Bodwin:1994jh}). In the language of
NRQCD, $\tilde{I}^{(2)}(\bm{p})$ is the perturbative one-loop
renormalization of the operators in Eq.~(\ref{psi20}) in leading order
in $\alpha_s$ and $v$.

Note that in $\tilde{I}^{(2)}(\bm{p})$ we
retain the binding energy of the heavy quarkonium
\begin{equation}
\epsilon_{\textrm{B}}=-\gamma_{\textrm{B}}^2/m.
\label{eps-b}
\end{equation}
We assign momenta such that the $Q$ and $\bar Q$ each carry half of the
binding energy in the rest frame of the heavy quarkonium.

$\Delta\psi^{(2)}(0)$ is given by 
\begin{equation}
\label{me-corr}
\Delta\psi^{(2)}(0)=
\int\frac{d^3p}{(2\pi)^3}\, 
\Delta\tilde{I}^{(2)}(\bm{p}) 
\,\widetilde{\psi}(\bm{p}),
\end{equation}
where 
\begin{equation}
\Delta\tilde{I}^{(2)}(\bm{p})=\tilde{I}_{\Lambda}^{(2)}(\bm{p})
-\tilde{I}_{\textrm{DR}}^{(2)}(\bm{p}).
\end{equation}
Writing out the expression for the diagram in Fig.~\ref{fig:Feynman} for 
the cases of a hard cutoff and dimensional regularization, we obtain
\begin{eqnarray}
\label{I0p}
\Delta\tilde{I}_{\textrm{B}}^{(2)}(\bm{p})&=&
4\pi\alpha_s C_F
\int \frac{d^4k}{(2\pi)^4}
(\bm{k}+\bm{p})^{2}
\frac{i}{\bm{k}^2}
\frac{i}{\left[k_0-\frac{\gamma_{\textrm{B}}^2}{2m}
                  -\frac{(\bm{k}+\bm{p})^2}{2m}
        +i\epsilon\right]}
\nonumber\\
&&\quad\quad\quad\times
\frac{i}{\left[-k_0-\frac{\gamma_{\textrm{B}}^2}{2m}
                   -\frac{(\bm{k}+\bm{p})^2}{2m}
        +i\epsilon\right]}
\left[
\frac{\Lambda^2}{(\bm{k}+\bm{p})^2+\Lambda^2}-1_{\rm DR}
\right].
\end{eqnarray}
The first term in the numerator brackets in Eq.~(\ref{I0p}) corresponds
to the hard cutoff, while the second term ``$1_{\rm DR}$'' corresponds to
dimensional regularization. The meaning of the ``$1_{\rm DR}$'' is that it is
unity unless it multiplies a scaleless integral (which vanishes in
dimensional regularization), in which case it is zero. The subscript 
``B'' in $\Delta\tilde{I}^{(2)}_{\rm B}(\bm{p})$ indicates that we have
retained the dependence on the binding energy $-\gamma_{\textrm{B}}^2/2m$.
Since the integral in Eq.~(\ref{I0p}) is dominated by large momenta, we
expect the final result to be insensitive to $\gamma_{\textrm{B}}$. However, in
discussions that occur later in this paper, it is illuminating to retain
the full $\gamma_{\textrm{B}}$ dependence.

The integral over the loop energy $k_0$ in Eq.~(\ref{I0p}) can be carried 
out by using the residue theorem. The result is
\begin{equation}
\label{dI}
\Delta\tilde{I}_{\textrm{B}}^{(2)}(\bm{p})=
4\pi \alpha_s C_F m
\int \frac{d^3{k}}{(2\pi)^3}
\frac{(\bm{k}+\bm{p})^{2}}
{\bm{k}^2 \left[(\bm{k}+\bm{p})^2+\gamma_{\textrm{B}}^2\right]}
\left[
\frac{\Lambda^2}{(\bm{k}+\bm{p})^2+\Lambda^2}
-1_{\rm DR}
\right].
\end{equation}
(Note that we also could have reached this expression directly from the
momentum-space Schr\"odinger equation.) In the second term in brackets
in Eq.~(\ref{dI}), we discard a scaleless integral of the form
\begin{equation}
\int \frac{d^3{k}}{(2\pi)^3}\frac{1_{\rm DR}}{\bm{k}^{2}}.
\label{scaless1}
\end{equation}
In the first term in brackets
in Eq.~(\ref{dI}), we rewrite the numerator $(\bm{k}+\bm{p})^2$ as 
$[(\bm{k}+\bm{p})^2+\gamma_{\textrm{B}}^2]-\gamma_{\textrm{B}}^2$ and partial-fraction 
the $-\gamma_{\textrm{B}}^2$ term. The result of these manipulations is
\begin{equation}
\Delta\tilde{I}^{(2)}_{\textrm{B}}(\bm{p})=
\frac{4\pi \alpha_s C_F m}{\Lambda^2-\gamma_{\textrm{B}}^2}
\int \frac{d^3{k}}{(2\pi)^3}
\frac{1}{\bm{k}^2}
\left[
\frac{\Lambda^4}{(\bm{k}+\bm{p})^2+\Lambda^2}
-
\frac{\gamma_{\textrm{B}}^4}{(\bm{k}+\bm{p})^2+\gamma_{\textrm{B}}^2}
\right].
\label{dIp}
\end{equation}
Evaluation of the integral in Eq.~(\ref{dIp}) is straightforward. The 
result is
\begin{equation}
\Delta
\tilde{I}^{(2)}_{\textrm{B}}(\bm{p})
   =\frac{\alpha_s C_F m}{|\bm{p}|(\Lambda^2-\gamma_{\textrm{B}}^2)}
  \left[ 
   \Lambda^4 \arctan\left(\frac{|\bm{p}|}{\Lambda}\right)
  -\gamma_{\textrm{B}}^4  
   \arctan\left(\frac{|\bm{p}|}{\gamma_{\textrm{B}}}\right)
  \right].
\label{Ipm-e}
\end{equation}
As expected, $\tilde{I}^{(2)}_{\textrm{B}}(\bm{p})$ is insensitive to 
$\gamma_{\textrm{B}}$. Neglecting terms of higher order 
in $\gamma_{\textrm{B}}^2/\Lambda^2$, 
we obtain
\begin{equation}
\Delta
\tilde{I}^{(2)}_{\textrm{NB}}(\bm{p})
   \equiv
\lim_{\gamma_{\textrm{B}}\to 0}
\Delta
\tilde{I}^{(2)}_{\textrm{B}}(\bm{p})
=
\frac{\alpha_s C_F m}{|\bm{p}|}
   \Lambda^2 \arctan\left(\frac{|\bm{p}|}{\Lambda}\right),
\label{Ipm-a}
\end{equation}
where the subscript ``NB'' indicates that we have neglected the binding
energy by dropping contributions of higher order in
$\gamma_{\textrm{B}}^2/\Lambda^2$. 
(Note that $\arctan(|\bm{p}|/\gamma_{\textrm{B}})$ is well behaved 
as $\gamma_{\textrm{B}}\to 0$ and bounded over the entire range of
its argument.) In our numerical analyses, we make use of
$\Delta\tilde{I}^{(2)}_{\textrm{NB}}(\bm{p})$, rather than
$\Delta\tilde{I}^{(2)}_{\textrm{B}}(\bm{p})$, and consistently neglect
the binding energy in short-distance (high-momentum) quantities. Of
course, the binding-energy dependence in the wave function is retained fully.

In our numerical analyses, we solve the Schr\"odinger equation
in coordinate space rather than in momentum space. The Fourier 
transformation of the coordinate-space wave function to momentum space 
involves an oscillating integrand and, hence, is difficult to evaluate 
numerically. Therefore, it is convenient to evaluate 
$\Delta\psi^{(2)}(0)$ in coordinate space:
\begin{equation}
\Delta\psi^{(2)}(0)
=\int d^3x\, \Delta I^{(2)}(\bm{x})\psi(\bm{x}),
\label{dpsi}
\end{equation}
where
\begin{equation}
\Delta I^{(2)}(\bm{x})=\int\frac{d^3{p}}{(2\pi)^3}e^{i\bm{p}\cdot\bm{x}} 
\Delta\tilde{I}^{(2)}(\bm{p})
\end{equation}
is the Fourier transform of $\Delta\tilde{I}^{(2)}(\bm{p})$. It is a 
simple matter to evaluate $\Delta I^{(2)}(\bm{x})$ analytically. The 
results are 
\begin{subequations}
\begin{eqnarray}
\label{Ix}
\Delta I^{(2)}_{\textrm{B}}(\bm{x})
&=&
\int\frac{d^3{p}}{(2\pi)^3}e^{i\bm{p}\cdot\bm{x}} 
\Delta\tilde{I}^{(2)}_{\textrm{B}}(\bm{p})
= \frac{ \alpha_s C_F m }{4\pi r^2(\Lambda^2-\gamma_{\textrm{B}}^2)}
\left(
\Lambda^4 e^{-\Lambda r}
-\gamma_{\textrm{B}}^4e^{-\gamma_{\textrm{B}} r}
\right),\label{Ixb}
\\
\Delta I^{(2)}_{\textrm{NB}}(\bm{x})
&=&
\int\frac{d^3{p}}{(2\pi)^3}e^{i\bm{p}\cdot\bm{x}}
\Delta\tilde{I}^{(2)}_{\textrm{NB}}(\bm{p})
= \frac{ \alpha_s C_F m \Lambda^2}{4\pi r^2}
e^{-\Lambda r}.\label{Ixnb}
\end{eqnarray}
\end{subequations}
Substituting $\Delta I^{(2)}(\bm{x})$ into Eq.~(\ref{dpsi}), we obtain
\begin{subequations}
\begin{eqnarray}
\label{dpsif}
\Delta \psi^{(2)}_{\textrm{B}}(0)
&=&
\frac{\alpha_s C_F m}{\Lambda^2-\gamma_{\textrm{B}}^2}
\int_0^\infty
\left(
\Lambda^4 e^{-\Lambda r}
-\gamma_{\textrm{B}}^4e^{-\gamma_{\textrm{B}} r}
\right)\frac{R(r)}{\sqrt{4\pi}}\,dr,
\\
\label{dpsif0}
\Delta \psi^{(2)}_{\textrm{NB}}(0)
&=&
\alpha_s C_F m\Lambda^2
\int_0^\infty
e^{-\Lambda r}
\frac{R(r)}{\sqrt{4\pi}}
\,dr.
\end{eqnarray}
\end{subequations}
Then, we obtain the dimensionally-regulated matrix element by computing
\begin{equation}
\psi^{(2)}_{\textrm{DR,}\Lambda}(0)=
\psi^{(2)}_\Lambda(0)
-\Delta\psi^{(2)}_{\textrm{NB}}(0)
\label{psiDRL}
\end{equation}
and taking the limit $\Lambda\to\infty$ in order to remove 
uncompensated power corrections that vanish as $1/\Lambda$:
\begin{equation}
\psi^{(2)}_{\textrm{DR}}(0)=
\lim_{\Lambda\to\infty}
\psi^{(2)}_{\textrm{DR,}\Lambda}(0).
\label{LIMpsiDRL}
\end{equation}

\section{Decomposition of $\bm{\Delta\psi^{(2)}(0)}$}
\label{decomposition}

It is illuminating to decompose $\Delta\psi^{(2)}(0)$ into products of
NRQCD matrix elements times short-distance coefficients.
In existing lattice calculations \cite{bks}, the difference between
matrix elements in hard-cutoff regularization and those in
dimensional regularization is expressed in terms of such a
decomposition. We decompose $\Delta\psi^{(2)}(0)$ by applying the
method of regions \cite{Beneke:1997zp} to the integration over the
wave-function momentum $\bm{p}$ in Eq.~(\ref{me-corr}). We take for
$\Delta \tilde{I}^{(2)}(\bm{p})$ the no-binding expression $\Delta
\tilde{I}^{(2)}_{\textrm{NB}}(\bm{p})$, which is given in
Eq.~(\ref{Ipm-a}).

First, we extract the leading term, which is 
obtained from the $|\bm{p}|^0$ part of 
$\Delta \tilde{I}^{(2)}_{\textrm{NB}}(\bm{p})$, namely, 
$\alpha_sC_Fm\Lambda$. It yields a contribution to 
$\Delta\psi^{(2)}(0)$ that is 
\begin{equation}
\Delta\psi_1^{(2)}(0)=\alpha_s C_F m \Lambda\int \frac{d^3p}{(2\pi)^3}
\widetilde{\psi}(\bm{p})
=\alpha_sC_Fm\Lambda\psi(0)=2\gamma_{\textrm{C}}\Lambda\psi(0).
\end{equation} 
We identify this as a one-loop short-distance coefficient times
the matrix element of leading order in $v$, namely, $\psi(0)$.
$\Delta\psi_1^{(2)}(0)$ is linearly divergent in the limit
$\Lambda\to\infty$ and cancels the linear divergence in the
hard-cutoff matrix element $\psi^{(2)}_{\Lambda}(0)$, resulting in
a matrix element $\psi^{(2)}_{\rm DR}(0)$, which, according to the NRQCD
$v$-scaling rules, is of order $(m v)^2\psi(0)$. The $v$ scaling of
this quantity is verified by the result of an analytic calculation of
$\psi^{(2)}_{\textrm{DR}}(0)$ for the case of a pure Coulomb potential
in Appendix~\ref{coulomb}.

Next we examine the remainder of $\Delta
\tilde{I}^{(2)}_{\textrm{NB}}(\bm{p})$, namely,
$\alpha_sC_Fm\Lambda[(\Lambda/|\bm{p}|)\,\arctan(|\bm{p}|/\Lambda)-1]$.
We decompose the loop integration over $\bm{p}$ into regions of small 
$|\bm{p}|$, in which $|\bm{p}|\ll \Lambda$, and large $|\bm{p}|$, 
in which $|\bm{p}|\sim \Lambda$. 

In the small-$|\bm{p}|$ region, we expand in powers of $|\bm{p}|$.
The resulting contributions, which we call $\Delta\psi_2^{(2)}(0)$, have 
the form
\begin{equation}
\Delta\psi_2^{(2)}(0)=\alpha_s C_F m \int_{\textrm{DR}} 
\frac{d^3p}{(2\pi)^3} \widetilde{\psi}(\bm{p}) \sum_{n=3}^\infty
c_n\frac{|\bm{p}|^{n-1}}{\Lambda^{n-2}},
\end{equation}
where $c_n$ is the coefficient of $(|\bm{p}|/\Lambda)^{n}$ in the
power-series expansion of $\arctan(|\bm{p}|/\Lambda)$, and the subscript 
$\textrm{DR}$ on the integral indicates that any UV divergence is to be 
regulated dimensionally.
$\Delta\psi_2^{(2)}(0)$ corresponds to one-loop short-distance
coefficients times dimensionally-regulated NRQCD matrix elements of
higher order in $v$. The contributions to $\Delta\psi_2^{(2)}(0)$ are of
order $\alpha_s m(mv)^{n-1}/\Lambda^{n-2}$ and, hence, are 
suppressed relative to $\psi^{(2)}_{\rm DR}(0)$ 
by at least one power of $m/\Lambda$ as $\Lambda\to \infty$.

In the large-$|\bm{p}|$ region, we approximate $\widetilde{\psi}(\bm{p})$ 
by its asymptotic form at large $|\bm{p}|$, which amounts to neglecting 
corrections of higher order in $\alpha_s$ and $mv/\Lambda$. The 
asymptotic form of $\widetilde{\psi}(\bm{p})$ is obtained by using the
Bethe-Salpeter equation to expose an explicit loop in the wave function,
as in the derivation of Eq.~(\ref{dI}). The result is
\begin{equation}
\widetilde{\psi}(\bm{p})\sim\int\frac{d^3k}{(2\pi)^3}\widetilde{\psi}(\bm{k})
\frac{8\pi\gamma_{\textrm{C}}}{(\bm{p}-\bm{k})^2\bm{p}^2}
\sim \frac{8\pi\gamma_{\textrm{C}}\psi(0)}{\bm{p}^4}
\equiv \widetilde{\psi}_{\rm asy}(\bm{p}).
\end{equation}
Then, we have for the contribution in the large-$|\bm{p}|$ region
\begin{eqnarray} 
\Delta\psi_3^{(2)}(0)
&=&\alpha_s C_F m \Lambda\int \frac{d^3p}{(2\pi)^3}
\widetilde{\psi}_{\rm asy}(\bm{p})
\left[\frac{\Lambda}{|\bm{p}|}\arctan\left(\frac{|\bm{p}|}{\Lambda}\right)
-1\right]\nonumber\\
&=&-(1/2)(\alpha_sC_Fm)^2\psi(0)=-2\gamma_{\textrm{C}}^2\psi(0).
\label{Delta-psi_3}
\end{eqnarray}
$\Delta\psi_3^{(2)}(0)$ has the interpretation of a two-loop short-distance 
coefficient times the lowest-order NRQCD matrix element. 

Although the contribution $\Delta\psi_3^{(2)}(0)$ is suppressed as
$\alpha_s/\Lambda$ relative to $\Delta\psi_1^{(2)}(0)$,
$\Delta\psi_3^{(2)}(0)$ can be important numerically because, in
$\psi^{(2)}_{\Lambda,{\rm DR}}(0)$, $\Delta\psi_1^{(2)}(0)$ is canceled
by contributions from $\psi^{(2)}_{\Lambda}(0)$, leaving a small
remainder. As we have remarked, $\psi^{(2)}_{\rm DR}(0)$ itself is
nominally of order $(m v)^2\psi(0)$. Therefore,
$\Delta\psi_3^{(2)}(0)$ is suppressed only as $\alpha_s^2/v^2$ relative
to $\psi^{(2)}_{\rm DR}(0)$. According to the $v$-scaling rules of NRQCD
\cite{Bodwin:1994jh}, $\alpha_s$ is of order $v^2$, and, hence, there is
a slight suppression. It turns out that, for the Cornell potential, the
numerical value of $\Delta\psi_3^{(2)}(0)$ is about $-40\%$ of
$\psi^{(2)}_{\textrm{DR}}(0)$. This suggests that two-loop corrections
to the short-distance coefficients could be important numerically in
converting results of lattice calculations of $\psi^{(2)}(0)$ to
continuum regularization. In the case of a pure Coulomb potential,
$\alpha_s$ is of order $v$, and, so, there is no suppression of
$\Delta\psi_3^{(2)}(0)$ relative to $\psi^{(2)}_{\rm DR}(0)$ at all. We
show in Appendix~\ref{coulomb}, $\psi^{(2)}_{\textrm{DR}}(0)$
satisfies the Gremm-Kapustin relation in the case of a pure Coulomb
potential, while $\psi^{(2)}_{\textrm{DR}}(0)-\Delta\psi_3^{(2)}(0)$ does
not.

\section{Direct method of calculation of 
$\bm{\psi}^{\bm{(2)}}_{\textrm{DR}}$\label{direct}(0)}
In the method of calculation that we have outlined, it is necessary to
take the limit of the quantity $\psi^{(2)}_{\textrm{DR,}\Lambda}(0)$ as
$\Lambda$ goes to infinity. $\psi^{(2)}_{\textrm{DR,}\Lambda}(0)$
consists of a difference of $\Lambda$-dependent terms that grow
approximately linearly with $\Lambda$. If one computes at a large enough
value of $\Lambda$ to be near the asymptotic value of
$\psi^{(2)}_{\textrm{DR,}\Lambda}(0)$, then there is a substantial
cancellation between these terms. For example, for the NLO parameters at
$\lambda=1.0$, the ratio
$\psi^{(2)}_{\Lambda}(0)/\psi^{(2)}_{\textrm{DR,}\Lambda}(0)$ takes on
the values 18, 34, and 50 when $\Lambda/m$ equals 10, 20, and 30,
respectively. In numerical calculations, the cancellation in
$\psi^{(2)}_{\textrm{DR,}\Lambda}(0)$ is a significant obstacle to
achieving good accuracy. Therefore, it is desirable to have a direct
method of computation of $\psi^{(2)}_{\textrm{DR}}(0)$ that does not
pass through an intermediate hard-cutoff step. In this section, we
present such a method. We also show how it leads to a simple formula for
the resummation of a class of corrections of higher order in $v$.

\subsection{Method}

We begin again by exposing one loop in the matrix element, as in 
Fig.~\ref{fig:Feynman}. Now, however, since we are computing the matrix 
element itself, not a difference of matrix elements for different 
regulators, the loop in Fig.~\ref{fig:Feynman} is not dominated by large 
momenta. Therefore, we must retain the complete Cornell potential in the 
corresponding expression. Repeating the steps that lead to 
Eq.~(\ref{dI}), but for the complete Cornell potential, and using 
Eq.~(\ref{1-loop-renorm}), we obtain for the dimensionally-regulated 
matrix element
\begin{equation}
\psi^{(2)}_{\textrm{DR}}(0)
=
-m\int\frac{d^3k}{(2\pi)^3}
\frac{\bm{k}^2 1_{\rm DR}}
     {\bm{k}^2+\gamma_{\textrm{B}}^2+i\epsilon}
\int\frac{d^3p}{(2\pi)^3}
\widetilde{V}(\bm{k}-\bm{p})\widetilde{\psi}(\bm{p}).
\label{psi2-direct-0}
\end{equation}
It turns out that, in the case of the Cornell potential, the binding 
energy is positive. Therefore, it is convenient to define a quantity 
$\widetilde{\gamma}_{\textrm{B}}$ as
\begin{equation}
\epsilon_{\textrm{B}}=
-\frac{ \gamma^2_{\textrm{B}} }{m}
=
\frac{ \widetilde{\gamma}^2_{\textrm{B}} }{m}>0.
\label{tilde-gamma-b}
\end{equation}
From the $i\epsilon$ prescription in Eq.~(\ref{psi2-direct-0}), we
see that we can analytically continue from $\gamma_{\textrm{B}}$
positive to $\gamma_{\textrm{B}}$ positive imaginary. Hence, we make the
identification $\gamma_{\textrm{B}}=i\widetilde{\gamma}_{\textrm{B}}$, 
where $\widetilde{\gamma}_{\textrm{B}}$ is real and positive.
Discarding the contribution of the scaleless integral 
\begin{equation}
\int\frac{d^3k}{(2\pi)^3}
\widetilde{V}(\bm{k}-\bm{p})1_{\rm DR}=
\int\frac{d^3k}{(2\pi)^3}                                        
\widetilde{V}(\bm{k})1_{\rm DR}
\label{scaleless}
\end{equation}
in Eq.~(\ref{psi2-direct-0}), we have
\begin{equation}
\psi^{(2)}_{\textrm{DR}}(0)
=
-m\widetilde{\gamma}^2_{\textrm{B}}
\int\frac{d^3k}{(2\pi)^3}
\int\frac{d^3p}{(2\pi)^3}
\frac{\widetilde{V}(\bm{k}-\bm{p})
\widetilde{\psi}(\bm{p})
}{\bm{k}^2-\widetilde{\gamma}^2_{\textrm{B}}+i\epsilon}
\label{psi2-direct-mom}.
\end{equation}
Now we note that $\widetilde{\psi}(\bm{p})$ satisfies the momentum-space 
Schr\"odinger equation 
\begin{equation}
(\bm{k}^2-\widetilde{\gamma}^2_{\textrm{B}})\widetilde{\psi}(\bm{k})
=-\int\frac{d^3p}{(2\pi)^3}
m\widetilde{V}(\bm{k}-\bm{p})\widetilde{\psi}(\bm{p}).
\label{schro-mom}
\end{equation}
Therefore, Eq.~(\ref{psi2-direct-mom}) can be written as
\begin{equation}
\psi^{(2)}_{\textrm{DR}}(0)
=\widetilde{\gamma}^2_{\textrm{B}}
\int\frac{d^3k}{(2\pi)^3}
\widetilde{\psi}(\bm{k})
=\widetilde{\gamma}^2_{\textrm{B}}\psi(0).
\label{psi2-direct-final}
\end{equation}

The Gremm-Kapustin relation \cite{Gremm:1997dq} follows from the NRQCD 
equations of motion at leading order nontrivial in $v$. It states that 
\begin{equation}
\langle v^2 \rangle=\frac{\psi^{(2)}(0)}{m_c^2\psi(0)}=
\frac{\epsilon_{\textrm{B}}}{m_c}+{\cal O}(v^4).
\label{gremm-kapustin}
\end{equation}
Using the definition of $\widetilde{\gamma}^2_{\textrm{B}}$ in
Eq.~(\ref{tilde-gamma-b}), we see that our result in
Eq.~(\ref{psi2-direct-final}) is precisely the Gremm-Kapustin relation
for the potential model with mass $m=m_c$. This is not surprising, since
a potential model can be derived from NRQCD at leading order in $v$
(Ref.~\cite{Brambilla:1999xf}) and, hence, implicitly respects the
NRQCD equations of motion at leading order in $v$.

\subsection{Resummation}

We note that the result in Eq.~(\ref{psi2-direct-final}) can easily be
generalized to the case of NRQCD operator matrix elements involving
additional powers of $\bm{\nabla}^2$, such as
\begin{subequations}
\begin{eqnarray}
\psi^{(2n)}(0)&\equiv&
\int\frac{d^3p}{(2\pi)^3}
\bm{p}^{2n} 
\widetilde{\psi}(\bm{p})
=\frac{1}{\sqrt{2N_c}}
\langle 0|\chi^\dagger\left(-\bm{\nabla}^2\right)^n\psi|H({}^1S_0)\rangle,
\\
\bm{\epsilon}\psi^{(2n)}(0)&\equiv&\bm{\epsilon}\int\frac{d^3p}{(2\pi)^3}
\bm{p}^{2n}
\widetilde{\psi}(\bm{p})
=\frac{1}{\sqrt{2N_c}}
\langle 0|\chi^\dagger\bm{\sigma}\left(-\bm{\nabla}^2\right)^n
\psi|H({}^3S_1)\rangle,
\end{eqnarray}
\end{subequations}
where $n$ is an integer. As we have remarked earlier, such operator matrix
elements are usually written in terms of the covariant derivative
$\bm{D}$ (Ref.~\cite{Bodwin:1994jh}), rather than $\bm{\nabla}$.
However, in the Coulomb gauge, the difference between $\bm{D}$ and
$\bm{\nabla}$ is suppressed as $v$ (Ref.~\cite{Bodwin:1994jh}).

Repeating the steps that lead to Eq.~(\ref{psi2-direct-0}), we obtain
\begin{equation}
\psi^{(2n)}_{\textrm{DR}}(0)
=
-m\int\frac{d^3k}{(2\pi)^3}
\frac{\bm{k}^{2n} 1_{\rm DR}}
     {\bm{k}^2-\widetilde{\gamma}_{\textrm{B}}^2+i\epsilon}
\int\frac{d^3p}{(2\pi)^3}
\widetilde{V}(\bm{k}-\bm{p})\widetilde{\psi}(\bm{p}).
\label{psi2n-direct-0}
\end{equation}

In Eq.~(\ref{psi2n-direct-0}), we write
\begin{equation}
\frac{\bm{k}^{2n}}{\bm{k}^2-\widetilde{\gamma}_{\textrm{B}}^2+i\epsilon}
=\bm{k}^{2n-2}+\widetilde{\gamma}_{\textrm{B}}^2\bm{k}^{2n-4}+\cdots
+\frac{\widetilde{\gamma}_{\textrm{B}}^{2n}}
{\bm{k}^2-\widetilde{\gamma}_{\textrm{B}}^2+i\epsilon}
\end{equation}
and discard scaleless integrals of the form
\begin{equation}
\int\frac{d^3k}{(2\pi)^3}
\widetilde{V}(\bm{k}-\bm{p})\bm{k}^{2n'}1_{\rm DR}
=\int\frac{d^3k}{(2\pi)^3}
\widetilde{V}(\bm{k})(\bm{k}+\bm{p})^{2n'}1_{\rm DR},
\end{equation}
where $n'$ is an integer. The result is
\begin{equation}
\psi^{(2n)}_{\textrm{DR}}(0)
=-m\widetilde{\gamma}^{2n}_{\textrm{B}}
\int\frac{d^3k}{(2\pi)^3}
\int\frac{d^3p}{(2\pi)^3}
\frac{\widetilde{V}(\bm{k}-\bm{p})
\widetilde{\psi}(\bm{p})
}{\bm{k}^2-\widetilde{\gamma}^2_{\textrm{B}}+i\epsilon}
=\widetilde{\gamma}^{2n}_{\textrm{B}}\psi(0),
\label{psi2n-final}
\end{equation}
where we have used the momentum-space Schr\"odinger equation
[Eq.~(\ref{schro-mom})].
The expression in Eq.~(\ref{psi2n-final}) can be thought of as a
generalized Gremm-Kapustin relation that holds in a potential model.

Now, suppose that there are contributions to a decay or production 
amplitude, computed in NRQCD, of the form 
\begin{equation}
A^{(2n)}=\frac{1}{n!}H^{(2n)}(0)\psi_{\textrm{DR}}^{(2n)}(0). 
\end{equation}
Here, $(1/n!)H^{(2n)}(0)$ 
is an NRQCD short-distance coefficient that is defined by 
\begin{equation}
H^{(2n)}(0)=\left. 
\left(\frac{\partial} {\partial \bm{p}^2}\right)^{n}
   H(\bm{p}^2)
\right|_{\bm{p}=0},
\end{equation}
where $H^{(2n)}(\bm{p}^2)$ is the hard-scattering amplitude for the 
process, evaluated in the quarkonium rest frame at $Q$ or $\bar Q$ 
momentum squared $\bm{p}^2$.
We can use Eq.~(\ref{psi2n-final}) to resum the contributions $A^{(2n)}$:
\begin{equation}
\sum_n A^{(2n)}=\sum_n \frac{1}{n!}H^{(2n)}(0)\psi_{\textrm{DR}}^{(2n)}(0)
=\sum_n \frac{1}{n!}H^{(2n)}(0)\widetilde{\gamma}^{2n}_{\textrm{B}}\psi(0)
=H(\widetilde{\gamma}^2_{\textrm{B}})\psi(0).
\end{equation}

\section{Numerical results and discussion \label{numerical}}
We now use the hard-cutoff method of Secs.~\ref{cutoff} and \ref{difference} 
and the direct method of Sec.~\ref{direct} to compute 
$\psi_{\textrm{DR}}^{(2)}(0)$.  

\begin{figure}
\begin{tabular}{cc}
\includegraphics[width=8.0cm]{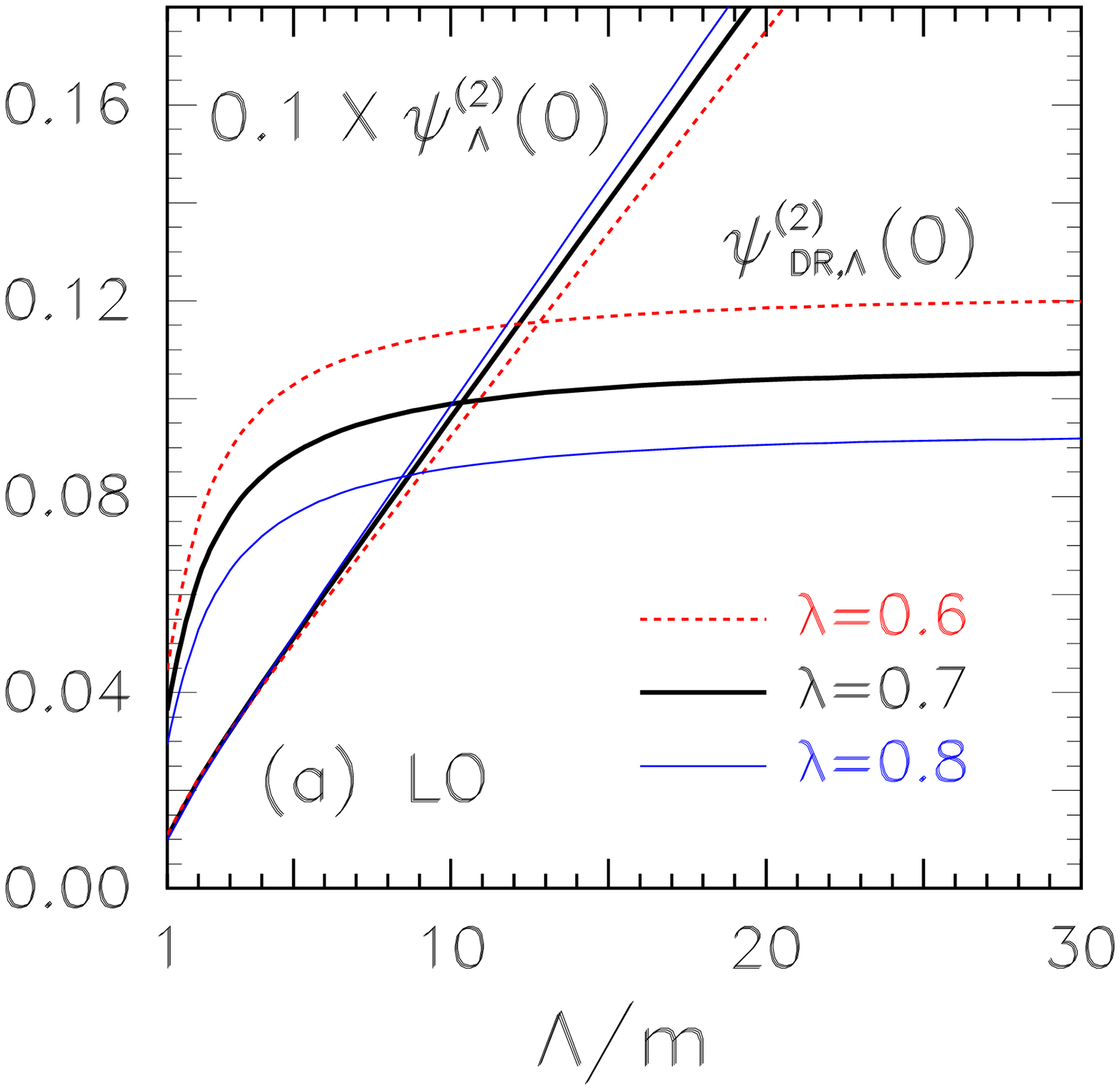}&
\includegraphics[width=8.0cm]{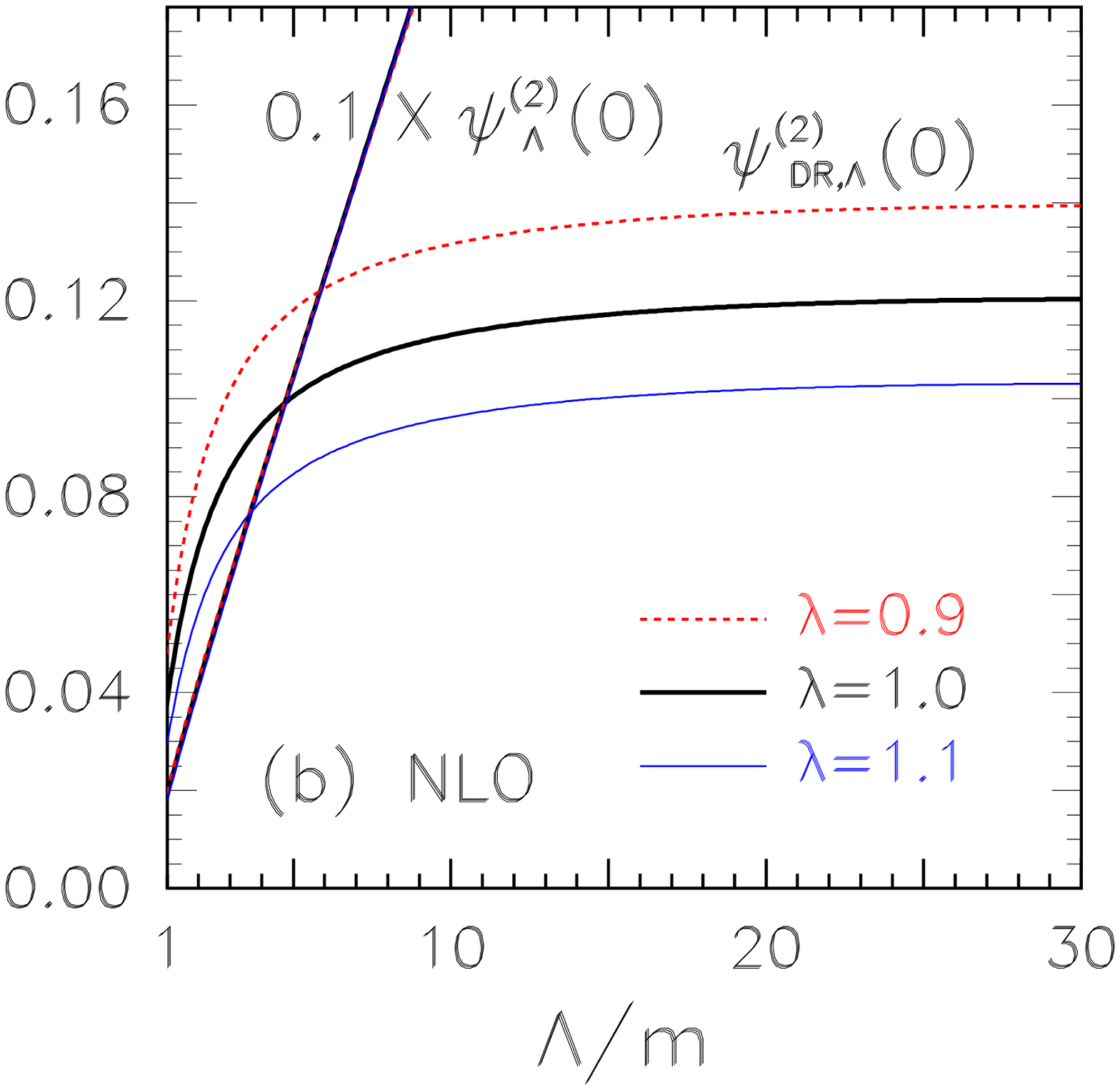}\\[-2.0ex]
\end{tabular}
\caption{\label{fig:me}
$\psi_\Lambda^{(2)}(0)$ and $\psi_{\textrm{DR,}\Lambda}^{(2)}(0)\equiv%
\psi^{(2)}_{\Lambda}(0)-\Delta\psi^{(2)}_{\textrm{NB}}(0)$ as a function
of $\Lambda/m$. The left figure corresponds to the LO
potential-model parameters of Table~\ref{table:ma-lo} for $\lambda=0.6$,
$0.7$, and $0.8$. The right figure corresponds to the NLO
potential-model parameters of Table~\ref{table:ma-nlo} for
$\lambda=0.9$, $1.0$, and $1.1$. In each figure, the three curves that
are nearly linear are $0.1\times \psi_\Lambda^{(2)}(0)$, and the three
curves that reach a plateau at large $\Lambda/m$ are
$\psi_{\textrm{DR,}\Lambda}^{(2)}(0)$.
}
\end{figure}

We begin with the hard-cutoff method. We substitute the $1S$-state
coordinate-space radial wave function that we obtain by integrating the
Schr\"odinger equation numerically into  Eqs.~(\ref{psi-lamr}) and
(\ref{dpsif0}) and carry out the integrations over $r$ numerically. We
then use Eq.~(\ref{psiDRL}) to compute
$\psi^{(2)}_{\textrm{DR,}\Lambda}(0)$. The results are shown in
Fig.~\ref{fig:me}. 
As can be seen, $\psi_\Lambda^{(2)}(0)$ grows nearly linearly with
$\Lambda$, as is expected from the linear ultraviolet divergence that it
contains. $\psi_{\textrm{DR,}\Lambda}^{(2)}(0)$ is the result of a
substantial cancellation between $\psi^{(2)}_{\Lambda}(0)$ and
$\Delta\psi^{(2)}_{\textrm{NB}}(0)$, at the level of about one part in
50 at $\Lambda/m=30$ for the NLO parameters. From Fig.~\ref{fig:me},
it is apparent that $\psi_{\textrm{DR},\Lambda}^{(2)}(0)$ reaches a
plateau at large $\Lambda$. $\psi_{\textrm{DR},\Lambda}^{(2)}(0)$
deviates from its asymptotic value by an amount that is of order
$1/\Lambda$. It is the value of $\psi_{\textrm{DR},\Lambda}^{(2)}(0)$ at
the plateau that corresponds to the dimensionally-regulated matrix
element [Eq.~(\ref{LIMpsiDRL})]. A fit of the LO-parameter results for
$\psi_{\textrm{DR,}\Lambda}^{(2)}(0)$ at $\lambda=0.6$, $0.7$, and $0.8$
to a constant plus a term that is proportional to $1/\Lambda$ yields
$\psi_{\textrm{DR},\Lambda}^{(2)}(0) =0.123$~GeV$^{7/2}$,
$0.108$~GeV$^{7/2}$, and $0.095$~GeV$^{7/2}$, respectively. A similar fit
to the NLO-parameter results at $\lambda=0.9$, $1.0$, and $1.1$ yields
$\psi_{\textrm{DR},\Lambda}^{(2)}(0) =0.144$~GeV$^{7/2}$, $0.124$~GeV$^{7/2}$,
and $0.107$~GeV$^{7/2}$, respectively.

Next we consider the direct method for calculating
$\psi_{\textrm{DR}}^{(2)}(0)$. Substituting the LO potential-model
parameters into Eq.~(\ref{psi2-direct-final}), we obtain
$\psi_{\textrm{DR}}^{(2)}(0)=0.124$~GeV$^{7/2}$, $0.109$~GeV$^{7/2}$,
and $0.096$~GeV$^{7/2}$ at $\lambda=0.6$, $0.7$, and $0.8$,
respectively. The results for the NLO-parameters are
$\psi_{\textrm{DR}}^{(2)}(0)=0.146$~GeV$^{7/2}$, $0.127$~GeV$^{7/2}$,
and $0.109$~GeV$^{7/2}$ at $\lambda=0.9$, $1.0$, and $1.1$,
respectively. These results for $\psi_{\textrm{DR}}^{(2)}(0)$ are
in good agreement with those from the extrapolation of
$\psi_{\textrm{DR,}\Lambda}^{(2)}(0)$ to $\Lambda=\infty$ in the
hard-cutoff method.

We consider the NLO value for $\psi(0)$ to be slightly more reliable
than the LO value. Therefore, we use the NLO parameters to compute our
central value for $\psi_{\textrm{DR}}^{(2)}(0)$. We use the LO
parameters to give an indication of the uncertainty in
$\psi_{\textrm{DR}}^{(2)}(0)$ that arises from the uncertainty in
$\psi(0)$. As we have already discussed in Sec.~\ref{model}, the
potential-model values of the string tension that derive from the
NLO parameters for $\lambda=1.0$ and $1.1$ span the range of lattice
values for the string tension. The potential-model values of $\alpha_s$
that derive from the NLO parameters for $\lambda=1.0$ 
and $1.1$ are somewhat larger than the lattice values for the fixed parameter
$\alpha_s$, but are compatible with the value of the running $\alpha_s$
at the scale of the heavy-quark momentum $m_cv$. We consider the lattice
value of the string tension to be more relevant than the lattice value
of the fixed parameter $\alpha_s$, since the latter does not take into
account the running of $\alpha_s$ in QCD. Therefore, we determine the
central value of $\psi_{\textrm{DR}}^{(2)}(0)$ by taking the average of
the values of $\psi_{\textrm{DR}}^{(2)}(0)$ for the NLO parameters at
$\lambda=1.0$ and $1.1$. We take the difference of these values as the
uncertainty that is attributable to the uncertainty in the
potential-model parameters. We take the difference between the average
of the values of $\psi_{\textrm{DR}}^{(2)}(0)$ for the NLO parameters at
$\lambda=1.0$ and $1.1$ and the average of the values of
$\psi_{\textrm{DR}}^{(2)}(0)$ for the LO parameters at $\lambda=0.7$ and
$0.8$ to be the uncertainty that is attributable to the
uncertainty in $\psi(0)$. We add these uncertainties in quadrature. We
include an additional $30\%$ uncertainty to account for the fact that
our potential model neglects terms of relative order $v^2\approx
0.3$. Using the direct-method values for $\psi_{\textrm{DR}}^{(2)}(0)$,
we obtain
\begin{equation}
\psi_{\textrm{DR}}^{(2)}(0)= 0.118\pm 0.024\pm 0.035~\hbox{GeV}^{7/2},
\label{psi2-final}
\end{equation}
where the second uncertainty arises from the $v^2$ error. This is the 
dominant source of error.

The effects of the uncertainty in the value of $\psi(0)$ tend to cancel
in the ratio $\psi_{\textrm{DR}}^{(2)}(0)/\psi(0)$. We compute this
ratio for the LO-parameter and NLO-parameter direct-method results and
use the same method for determining the central value and the
uncertainties that we described above for $\psi_{\textrm{DR}}^{(2)}(0)$.
The result is
\begin{equation}
\psi_{\textrm{DR}}^{(2)}(0)/\psi(0)=m_c^2v^2=0.50\pm 0.09\pm 0.15~{\rm 
GeV}^2,
\label{final-ratio}
\end{equation}
where, again, the second uncertainty arises from the $v^2$ error.
Taking $m_c=1.4$~GeV in Eq.~(\ref{final-ratio}),
we have $\langle v^2 \rangle=0.25\pm 0.05\pm 0.08 $, which is in good
agreement with expectations from the NRQCD $v$-scaling rules.

We can now address the numerical importance of the two-loop 
contribution to $\Delta\psi^{(2)}(0)$
that is given in Eq.~(\ref{Delta-psi_3}).
Taking the average of the values of $\gamma_{\textrm{C}}$ for
$\lambda=1.0$ and $1.1$ in Table~\ref{table:ma-nlo} and using the NLO
value $\psi(0)=0.23629$~GeV$^{3/2}$, we find that
$\Delta\psi_3^{(2)}(0)=-0.048$~GeV$^{7/2}$. This is about $-40\%$ of the
value of $\psi_{\textrm{DR}}^{(2)}(0)$ in Eq.~(\ref{psi2-final}).

The $v$-scaling rules of NRQCD state that the hard-cutoff matrix
element $\psi_{\Lambda}^{(2)}(0)$, evaluated at $\Lambda\sim
mv$, should be of order $m_c^2v^2\psi(0)$. Taking $m_c=1.4$~GeV,
$v^2=0.3$, and $\psi(0)=0.23629$~GeV$^{3/2}$, we obtain
$m_c^2v^2\psi(0)\approx 0.14$~GeV$^{7/2}$, which is in reasonably
good agreement with the value of $\psi_{\Lambda}^{(2)}(0)$ at
$\Lambda=m_c v\approx m$ for the NLO results at $\lambda=1.0$ and 
$1.1$ in Fig.~\ref{fig:me}.

In lattice determinations of $\psi_{\textrm{DR},\Lambda}^{(2)}(0)$,
the lattice ultraviolet cutoff, which is of order $\pi$ divided by the
lattice spacing, corresponds approximately to the hard cutoff $\Lambda$.
Existing lattice determinations of $\psi_{\textrm{DR,}\Lambda}^{(2)}(0)$
(Ref.~\cite{bks}) have been carried out in the vicinity of
$\Lambda=m_c$.  This value of $\Lambda$ is at the boundary of the region
in which asymptotic freedom allows one to evaluate quantities in QCD in
perturbation theory. However, as can be seen from Fig.~\ref{fig:me},
$\Lambda=m_c$ is far from the region in which
$\psi_{\textrm{DR,}\Lambda}^{(2)}(0)$ approaches its asymptotic value.
Apparently, at $\Lambda=m_c$, power corrections of order $m_cv/\Lambda$
are still important. Therefore, we expect that the lattice
determinations of $\psi_{\textrm{DR}}^{(2)}(0)$ contain large
$1/\Lambda$ errors.

\section{Summary \label{conclusions}}

In this paper we have presented two methods for calculating NRQCD matrix
elements that are proportional to $\psi^{(2)}(0)$, the negative of the
second derivative of the wave function at the origin. These matrix
elements enter into the relativistic corrections for the decay and
production of heavy-quarkonium states. The matrix elements are linearly
ultraviolet divergent, and, hence, they must be regulated. We compute
the matrix elements in dimensional regularization, since that is the
regularization that is used most commonly in phenomenology.

One method that we have presented makes use of a hard-cutoff regulator
as an intermediate step, then employs perturbation theory to compute the
difference between the hard-cutoff matrix element and a
dimensionally-regulated matrix element. This method is quite general, in
that it requires knowledge only of the $Q\bar Q$ wave function in the
quarkonium state. In principle, it could be applied to wave functions
that are determined directly by lattice methods. It involves computing
cancelling quantities in the limit in which the hard cutoff is taken to
infinity. This method has the disadvantage that the limiting procedure and
cancellation make it difficult to achieve adequate numerical accuracy.
In the case of a pure Coulomb potential, we have obtained analytic
expressions for the quantities that enter into the hard-cutoff method.
The result for the matrix elements agrees with the Gremm-Kapustin relation 
in that case.

The second method that we have presented involves a computation of the
matrix elements directly in dimensional regularization. The method is
specific to potential models. It yields the Gremm-Kapustin
relation for the matrix elements. This is not surprising, since the
Gremm-Kapustin relation is based on the NRQCD equations of motion at
leading-order in $v$, and the potential model respects those equations of
motion. This method is easily generalized to the 
computation of matrix elements that are proportional to higher powers of
$\bm{\nabla}^2$ acting on the wave function at the origin. We have used
the results for such matrix elements to write a formula that resums 
certain contributions of higher order in $v$ to NRQCD amplitudes.

We have used both the direct method and the hard-cutoff method to
evaluate the dimensionally-regulated quantity
$\psi_{\textrm{DR}}^{(2)}(0)$ for the $J/\psi$ (or the $\eta_c$) in the
Cornell potential model. Since the Cornell potential model contains no
spin dependence, we do not distinguish between the $J/\psi$ and $\eta_c$
matrix elements. If the potential itself were exact, then the
potential model would reproduce QCD up to corrections of relative order
$v^2$. Existing lattice measurements of the static $Q\bar Q$ potentials
yield values for the string tension. In order to estimate errors from
our choice of potential, we have carried out computations for sets of
Cornell-potential parameters that bracket the lattice values of the
string tension.

The two methods for computing $\psi_{\textrm{DR}}^{(2)}(0)$ yield
results that agree well numerically. Our final result, including
estimates of uncertainties, is given in Eq.~(\ref{psi2-final}). The
first error estimate arises from the uncertainty in the
potential-model parameters and from the uncertainty in the value of the
wave function at the origin that is obtained from the leptonic width of
the $J/\psi$. The second error estimate accounts for a $30\%$
uncertainty that arises from the fact that we have neglected corrections
of relative order $v^2$. This is the dominant source of error. The
effects of the uncertainty in the wave function at the origin tend to
cancel in the ratio of matrix elements
$\psi^{(2)}_{\textrm{DR}}(0)/\psi(0)$ in Eq.~(\ref{final-ratio}). From
the ratio in Eq.~(\ref{final-ratio}), we estimate that $\langle
v^2\rangle=0.25\pm 0.05\pm 0.08$, which is in good agreement with the
$v$-scaling rules of NRQCD.

From our analysis, it is clear that, in the hard-cutoff method, there
are large corrections of order $m_cv/\Lambda$, even at cutoffs
$\Lambda\approx m_c$. This implies that existing lattice computations of
$\psi^{(2)}_{\textrm{DR}}(0)$ (Ref.~\cite{bks}) contain large errors
that arise from such power corrections. 

We also have identified a large contribution to the difference between
the dimensionally-regulated matrix element $\psi^{(2)}_{\textrm{DR}}(0)$
and the hard-cutoff matrix element $\psi^{(2)}_{\Lambda}(0)$ that arises
from a two-loop contribution to a particular short-distance coefficient
[Eq.~(\ref{Delta-psi_3})]. That contribution is suppressed only as
$\alpha_s^2/v^2$ relative to $\psi^{(2)}_{\textrm{DR}}(0)$ and is about
$-40\%$ of $\psi^{(2)}_{\textrm{DR}}(0)$ in the case of the $J/\psi$.
Therefore, the two-loop contribution to this short-distance coefficient
may be important numerically when one converts a lattice-regulated value
of $\psi^{(2)}(0)$ to a dimensionally-regulated value.

\begin{acknowledgments}
We wish to thank Eric Braaten for useful discussions.
The research of G.T.B. in the High Energy Physics Division at
Argonne National Laboratory is supported by the U.~S.~Department of
Energy, Division of High Energy Physics, under Contract No.W-31-109-ENG-38.
The work of J.L. is supported by the Korea Research Foundation
under Grant No. KRF-2004-015-C00092.
The work of D.K. is supported in part by the Seoul Science 
Fellowship of Seoul Metropolitan Government.
J.L. thanks the High Energy Theory Group at Argonne National Laboratory
for its hospitality.
\end{acknowledgments}
\appendix
\section{Test of the method with a Coulomb wave function \label{coulomb}}

In this appendix, we apply our methods for calculating
$\psi_{\textrm{DR}}^{(2)}(0)$ to the case of the wave function for 
a pure Coulomb potential. In this case, we can evaluate all of the
relevant expressions analytically. Ultimately, we compare our results
with the prediction of the Gremm-Kapustin relation.

Coulomb wave functions in coordinate space and momentum space are 
\begin{subequations}
\label{coulomb-wavefunction}
\begin{eqnarray}
\psi_{\textrm{C}}(\bm{x})&=&\psi_{\textrm{C}}(0)e^{-\gamma_{\textrm{C}}r},
\label{coulomb-wavefunction-coor}\\
\widetilde{\psi}_{\textrm{C}}(\bm{p})&=&
\int d^3 x e^{-i\bm{p}\cdot\bm{x}}\psi_{\textrm{C}}(\bm{x})
=8\pi\psi_{\textrm{C}}(0)
\frac{\gamma_{\textrm{C}}}{(\bm{p}^2+\gamma^2_{\textrm{C}})^2},
\label{coulomb-wavefunction-mom}
\end{eqnarray}
\end{subequations}
where $\gamma_{\textrm{C}}$ is defined in Eq.~(\ref{gamma-C}). The 
corresponding binding energy is 
\begin{equation}
\epsilon_{\textrm{B}}=-\gamma^2_{\textrm{C}}/m.
\label{coul-binding}
\end{equation}
It then follows from Eq.~(\ref{eps-b}) that
\begin{equation}
\gamma_{\textrm{B}}=\gamma_{\textrm{C}}.
\end{equation}

We first carry out the calculation of $\psi_{\textrm{DR}}^{(2)}(0)$ in
the hard-cutoff method. Substituting the Coulomb wave function into
either Eq.~(\ref{psi-lam}) or Eq.~(\ref{psi-lamr}), we obtain the
hard-cutoff matrix element $\psi^{(2)}_{\Lambda}(0)$:
\begin{equation}
\psi^{(2)}_{\Lambda}(0)
=\Lambda^2 \psi_{\textrm{C}}(0)
\left[ 1-\frac{\Lambda^2}{(\Lambda+\gamma_{\textrm{C}})^2} \right].
\end{equation}
The calculation of $\psi^{(2)}_{\Lambda}(0)$ can be carried out by using
Eqs.~(\ref{dpsi}) and (\ref{Ix}). For the difference between the 
hard-cutoff and dimensionally-regulated matrix elements, we have
\begin{equation}
\Delta\psi^{(2)}_{\textrm{B}}(0)=
\psi_{\textrm{C}}(0)
\frac{2\gamma_{\textrm{C}}}{\Lambda^2-\gamma^2_{\textrm{B}}}
\left(\frac{\Lambda^4}{\Lambda+\gamma_{\textrm{C}}}
      -\frac{\gamma^4_{\textrm{B}}}
            {\gamma_{\textrm{B}}+\gamma_{\textrm{C}}} \right),
\label{dpsi2-C}
\end{equation}
where we have used the definition (\ref{gamma-C}) to replace the
prefactor $\alpha_s C_F m$ with $2\gamma_{\textrm{C}}$. In the limit 
$\gamma_{\textrm{B}}\to 0$, Eq.~(\ref{dpsi2-C}) becomes
\begin{equation}
\label{dpsi2-C-approx}
\Delta\psi^{(2)}_{\textrm{NB}}(0)
=
\Lambda^2\psi_{\textrm{C}}(0)
\frac{2\gamma_{\textrm{C}}}
     {\Lambda+\gamma_{\textrm{C}}}.
\end{equation}
It follows from Eq.~(\ref{psiDRL}) that
$\psi_{\textrm{DR,}\Lambda}^{(2)}(0)$ is given by
\begin{equation}
\psi_{\textrm{DR,}\Lambda}^{(2)}(0)
=
\psi_{\Lambda}^{(2)}(0)
-
\Delta\psi_{\textrm{NB}}^{(2)}(0)
=-\gamma^2_{\textrm{C}}\psi_{\textrm{C}}(0)
\times \left(\frac{\Lambda}{\Lambda+\gamma_{\textrm{C}}}\right)^2.
\label{psi2-dr-LO}
\end{equation}
Taking the limit $\Lambda \to \infty$, 
we obtain the dimensionally-regulated matrix element
$\psi_{\textrm{DR}}^{(2)}(0)$:
\begin{equation}
\psi_{\textrm{DR}}^{(2)}(0)
=\lim_{\Lambda\to \infty}\psi_{\textrm{DR,}\Lambda}^{(2)}(0)
=-\gamma^2_{\textrm{C}}\psi_{\textrm{C}}(0).
\label{psi2-COULOMB-ans}
\end{equation}
We note that the deviations from the asymptotic value go as
$\gamma_{\textrm{C}}/\Lambda\sim mv/\Lambda$ as $\Lambda\to\infty$.
Incidentally, had we retained the effects of the binding energy
by using $\Delta\psi_{\textrm{B}}^{(2)}(0)$ instead of 
$\Delta\psi_{\textrm{NB}}^{(2)}(0)$, then we would have obtained
\begin{equation}
\psi_{\textrm{DR,}\Lambda}^{(2)}(0)=\psi_{\Lambda}^{(2)}(0)
-
\Delta\psi_{\textrm{B}}^{(2)}(0)
=-\gamma^2_{\textrm{C}}\psi_{\textrm{C}}(0),
\end{equation}
where we have made use of the Coulomb-potential relation
$\gamma_{\textrm{B}}=\gamma_{\textrm{C}}$. In this case, the $\Lambda$
dependence would have vanished. However, this simplification is a
special property of the pure Coulomb case that arises from the fact that
the Coulomb-gluon exchange in $\Delta\psi_{\textrm{B}}^{(2)}(0)$ is, in
this special case, an interaction of the complete potential.

The result in Eq.~(\ref{psi2-COULOMB-ans}) agrees with the result from
the direct method in Eq.~(\ref{psi2-direct-final}) and with the
Gremm-Kapustin relation (\ref{gremm-kapustin}). We note that, had we
failed to include the two-loop correction to $\Delta\psi^{(2)}(0)$ in
Eq.~(\ref{Delta-psi_3}), then we would have obtained
$\psi^{(2)}_{\textrm{DR}}(0)=-3\gamma^2_{\textrm{C}}\psi_{\textrm{C}}(0)$,
which does not agree with the Gremm-Kapustin relation.

\section{Integrals}\label{integrals}
Here we record some integrals that are useful in deriving 
the expressions in this paper.
\subsection{Loop integrals}
\begin{subequations}
\begin{eqnarray}
\int \frac{d^3k}{(2\pi)^3} 
\frac{1}{(\bm{k}-\bm{p})^2(\bm{k}^2+\Lambda^2+i\epsilon)}
&=& 
\frac{i}{8\pi|\bm{p}|}
\log
\left(
\frac{\Lambda-i|\bm{p}|}{\Lambda+i|\bm{p}|}
\right)\nonumber\\
&=&
\frac{1}{4\pi|\bm{p}|}\arctan\left(\frac{|\bm{p}|}{\Lambda}\right),
\label{loop-1}
\\
\int \frac{d^3k}{(2\pi)^3} 
\frac{1}{(\bm{k}-\bm{p})^2(\bm{k}^2+\Lambda^2+i\epsilon)^2}
&=& \frac{1}{8\pi\Lambda(\bm{p}^2+\Lambda^2)},
\label{loop-2}
\\
\int \frac{d^3k}{(2\pi)^3} 
\frac{ \arctan\left(|\bm{k}|/\Lambda\right) }
     {|\bm{k}|(\bm{k}^2+\Lambda^2+i\epsilon)^2}
&=& \frac{1}{16\pi\Lambda^2}.
\label{loop-3}
\end{eqnarray}
\end{subequations}
Analytically continuing $\Lambda$ to $i\Lambda$, we obtain the following 
formulas:
\begin{subequations}
\begin{eqnarray}
\int \frac{d^3k}{(2\pi)^3}
\frac{1}{(\bm{k}-\bm{p})^2(\bm{k}^2-\Lambda^2+i\epsilon)}
&=& 
\frac{1}{4\pi |\bm{p}|}\arctan\left(\frac{|\bm{p}|}{i\Lambda}\right)
\nonumber\\
&=&-\frac{i}{4\pi |\bm{p}|}
\tanh^{-1}\left(\frac{|\bm{p}|}{\Lambda}\right),
\label{loop-1m}
\\
\int \frac{d^3k}{(2\pi)^3}
\frac{1}{(\bm{k}-\bm{p})^2(\bm{k}^2-\Lambda^2+i\epsilon)^2}
&=& \frac{-i}{8\pi\Lambda(\bm{p}^2-\Lambda^2)},
\label{loop-3m}
\\
\int \frac{d^3k}{(2\pi)^3}
\frac{ \arctan\left(-i|\bm{k}|/\Lambda\right) }
     {|\bm{k}| (\bm{k}^2-\Lambda^2+i\epsilon)^2}
&=& -\frac{1}{16\pi\Lambda^2}.
\label{loop-4m}
\end{eqnarray}
\end{subequations}
\subsection{Fourier Transformation}
\begin{subequations}
\begin{eqnarray}
\int \frac{d^3p}{(2\pi)^3}\, e^{i\bm{p}\cdot\bm{x}}
\frac{\arctan(|\bm{p}|/\Lambda)}{|\bm{p}|}
&=& \frac{e^{-\Lambda r}}{4\pi r^2},
\\
\int \frac{d^3p}{(2\pi)^3}\, e^{i\bm{p}\cdot\bm{x}}
\frac{1}{(\bm{p}^2+\Lambda^2)}
&=& \frac{e^{-\Lambda r}}{4\pi r},
\\
\int \frac{d^3p}{(2\pi)^3}\, e^{i\bm{p}\cdot\bm{x}}
\frac{1}{(\bm{p}^2+\Lambda^2)^2}
&=& \frac{e^{-\Lambda r}}{8\pi\Lambda},
\end{eqnarray}
\end{subequations}
where $r\equiv|\bm{x}|$.


\end{document}